\documentclass[
preprint,
preprintnumbers,
 amsmath,amssymb,
 aps,
]{revtex4-2}

\usepackage{graphicx}
\usepackage{dcolumn}
\usepackage{bm}
\usepackage[colorlinks=true,linkcolor=blue,citecolor=blue,urlcolor=blue]{hyperref}
\usepackage[nameinlink]{cleveref}
\usepackage{subfigure}
\usepackage{svg}
\begin{document}

\preprint{APS/123-QED}

\title{Effects of symmetry and hydrodynamics on the cohesion of groups of swimmers}

\author{Mohamed Niged Mabrouk}
\email{mmabrouk@uh.edu}
\author{Daniel Floryan}%
\email{dfloryan@uh.edu}
\affiliation{%
 Department of Mechanical and Aerospace Engineering, University of Houston, Houston, TX 77204, USA
}%

\date{\today}

\begin{abstract}
When groups of inertial swimmers move together, hydrodynamic interactions play a key role in shaping their collective dynamics, including the cohesion of the group. To explore how these interactions influence group cohesion, we develop a three-dimensional, inviscid, far-field model of a swimmer. Focusing on symmetric triangular, diamond, and circular group arrangements, we investigate whether passive hydrodynamics alone can promote cohesive behavior, and what role symmetry of the group plays. While small symmetric (and even asymmetric) groups can be cohesive, larger groups typically are not, instead breaking apart into smaller, self-organized subgroups that are cohesive. Notably, we discover circular arrangements of swimmers that chase each other around a circle, resembling the milling behavior of natural fish schools; we call this hydrodynamic milling. Hydrodynamic milling is cohesive in the sense that it is a fixed point of a particular Poincar\'e map, but it is unstable, especially to asymmetric perturbations. Our findings suggest that while passive hydrodynamics alone cannot sustain large-scale cohesion indefinitely, controlling interactions between subgroups, or controlling the behavior of only the periphery of a large group, could potentially enable stable collective behavior with minimal active input.
\end{abstract}

\maketitle

\section{\label{sec1:intro}Introduction}
In nature, collective behavior is widely observed across animal species~\cite{vicsek2012collective}, where local interactions among individuals give rise to emergent group-level patterns and dynamics. Such behavior is prevalent in groups ranging from insects and fish to birds and mammals, and is critical for enhancing survival and reproductive success \cite{parrish1997animal,couzin2005effective, landeau1986oddity, hemelrijk2015increased}. Collective coordination enables animals to achieve reciprocal benefits, such as more efficient predator avoidance \cite{major1978predator}, reduced individual predation risk, and improved foraging and navigation capabilities \cite{colvert2017local, pitcher1982fish}. Researchers across disciplines like robotics and engineering \cite{dorigo2020reflections, gibouin2018study, gazzola2016learning} have been drawn to this topic to engineer better swarm robots that work together in a coordinated manner and match nature's design principles. 

Though obvious, it is underappreciated that in order to reap the benefits of collective motion, the collective must remain intact, or cohesive. Maintaining a cohesive group is complicated by the presence of an environment that the individuals must contend with, such as the aerial environments of birds or aquatic environments of fish. Not only do such environments hold the potential to exert exogenous forcings on the individuals, such as when a bird encounters a gust, or when a group of fish encounters a turbulent flow~\cite{zhang2024collective}, but they also couple the individuals to each other through fluid-mediated interactions. In this work, we focus on the hydrodynamic interactions between swimmers---which are expected to play an outsized role in the collective motion of swimmers~\cite{ko2023role}---specifically on how they affect group cohesion. 

\citeauthor{weihs1973hydromechanics}' work \cite{weihs1973hydromechanics} was the first to link performance improvement in swimmers with hydrodynamic interactions. He argued that a diamond school arrangement is optimal because it creates constructive vortex interactions, allowing trailing fish to gain higher velocity without extra energy expenditure by benefiting from leaders' vortices. Subsequent research has focused on identifying the specific hydrodynamic conditions that enable energy savings through collective swimming formations. \citeauthor{ashraf2017simple} showed that at elevated swimming velocities, fish schools were observed to adopt lateral alignment patterns, with hydrodynamic analysis indicating this configuration reduces energetic costs during sustained high-speed movement~\cite{ashraf2017simple}. In another study, the authors showed that two fish swimming side by side with a difference in phase create a wall effect that reduces energy expenditure~\cite{zhang2024energy}.
Recent work by \citeauthor{heydari2024mapping} demonstrated that in larger groups, inline formations unevenly distribute hydrodynamic benefits, with trailing individuals gaining greater energy savings, up to a critical group size~\cite{heydari2024mapping}. In contrast, side-by-side formations were shown to provide equitable energy savings across all group members while maintaining stable cohesion at any scale. Schooling can evidently provide hydrodynamic benefits when the appropriate group structure is maintained. This motivates our investigation of freely swimming collectives and the dynamics governing their motion, which depend on social interactions, active decision-making, and passive hydrodynamic interactions. 

Behavioral models, such as the Vicsek alignment model~\cite{vicsek1995novel}, are widely used to model how individuals interact in a group. They offer a computationally efficient framework for simulating collective behavior in sizable groups. These models and their variants typically incorporate zones of attraction, avoidance, and alignment to describe how individual interactions depend on spatial distribution. 
The Vicsek model reproduces three stereotypical behaviors observed in fish schools: swarming, milling, and schooling~\cite{couzin2002collective, calovi2014swarming}. These behaviors are distinguished by varying degrees of polarization and rotational order, providing insights into the mechanisms underlying collective intelligence in fish groups. \citeauthor{tunstrom2013collective} further demonstrated that both group size and domain boundaries significantly influence the emergent dynamics of schools~\cite{tunstrom2013collective}. 

Although schooling dynamics primarily depend on social and sensory mechanisms~\cite{couzin2002collective, lafoux2023illuminance, pavlov2000patterns}, hydrodynamic effects significantly influence both the interactions and their collective outcomes~\cite{filella2018model}. Recent studies have combined behavioral models with hydrodynamic interactions to examine collective behavior in fish groups~\cite{filella2018model, zhou2025effect, huang2024collective}. These modeling approaches capture behavioral transitions similar to those observed in the Vicsek model while also considering boundary effects, group size, social interactions between individuals, and flow-mediated interactions. The aforementioned studies focus on two-dimensional schools with two-dimensional hydrodynamic interactions. While the precise effects of three-dimensionality on group cohesion and stability are unknown, they undoubtedly differ from the two-dimensional case since individual swimmers would not be constrained to planar configurations, and the strength and structure of three-dimensional hydrodynamic interactions differ from two-dimensional interactions. 

In previous work, we took a step towards understanding the effects of three-dimensionality by examining the fundamental three-dimensional hydrodynamic interactions between a pair of swimmers~\cite{mabrouk2024group}. Pairwise interactions serve as a building block for interactions between a larger number of individuals. We found that passive hydrodynamic interactions alone can lead to cohesive pairs of swimmers when their initial configurations are sufficiently well aligned. 

In this work, we examine emergent group dynamics arising from three-dimensional hydrodynamic interactions in larger swimmer populations, using the model developed in~\cite{mabrouk2024group}. The reduced-order modeling approach accounts for essential hydrodynamic coupling between swimmers. Social and other interactions are not accounted for, allowing us to probe purely hydrodynamic effects and examine to what degree cohesion can emerge from passive hydrodynamic interactions. The study focuses on initially symmetric configurations to determine whether these symmetries persist in the resulting dynamics---a characteristic that can influence group cohesion and promote stable collective motion.

The paper is structured as follows. Section~\ref{sec2:Model} reviews the modeling framework and derives the governing dynamical system. Following that, the free-swimming dynamics of several fundamental configurations of swimmers are explored: triangular configurations in Section~\ref{sec:tri}, diamond configurations in Section~\ref{sec:diamond}, and circular configurations in Section~\ref{sec4:Milling}. A summary and concluding remarks are presented in Section~\ref{sec6:Conclusion}. 
\section{\label{sec2:Model}Model Formulation}
\begin{figure}
    \centering
    \includegraphics[width=0.5\textwidth]{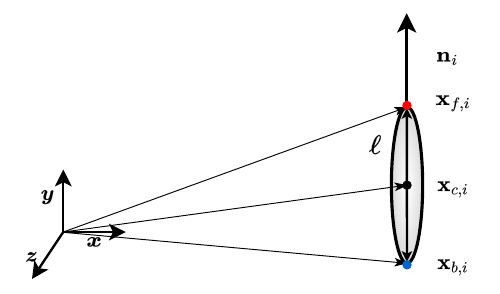}
    \caption{\label{fig3:SingleSwimmer} Representation of a swimmer modeled as source dipole.}     
\end{figure}

Each swimmer is modeled as a source-sink pair of equal strength separated by a fixed distance~$\ell$ (see Figure~\ref{fig3:SingleSwimmer}), producing a dipole flow at large distances. This is inspired by previous work on two-dimensional point-vortex dipoles~\cite{tchieu2012finite, tsang2013dipole}, but we emphasize that the present model is three-dimensional. Physically, the source represents the fore of the swimmer, where fluid is expelled in all directions as the swimmer advances, while the sink represents the rear, where fluid is drawn in to fill the void left behind. This model is based on the fact that the potential flow induced by a body of constant volume can be expressed as a multipole expansion in which the leading-order term is a dipole \cite[Ch. 4.7]{eldredge2019mathematical}. 
Outside vortical regions and away from boundary layers, the flow is irrotational, justifying the use of a potential flow model. The dipole term in the multipole expansion is a universal feature of swimmers, independent of their specific shape and kinematics---details that are instead encoded in higher-order terms.
Our model provides a computationally tractable way to calculate the flow induced by a swimmer, with broad applicability to inertial swimmers, regardless of their shape characteristics, enabling the study of inviscid mechanisms in groups of swimmers.

In isolation, our model swimmer moves at a speed $U = \frac{\sigma}{4 \pi \ell^2} $ along the direction of its body, where $\sigma \geq 0$ is the volumetric flow rate of the source/sink. We call $U$ the self-propelled speed. When more than one swimmer is present, they mutually affect each other's motions through the velocity fields they induce; that is, through their hydrodynamic interactions.

To determine the hydrodynamic interactions, we treat each swimmer as a dumbbell comprising two beads connected by a rigid rod, the beads respectively co-located with the source and sink. Such dumbbell representations are common in the microswimmer literature \cite{hernandez2005transport, hernandez2007fast, underhill2008diffusion, hernandez2009dynamics, ryan2011viscosity}. In this representation, all the hydrodynamic forces experienced by the swimmer are concentrated on the beads, and the rigid rod serves to keep the length of the dumbbell fixed. Using the results developed in~\cite{auton1988force} for the force on a sphere in an unsteady inviscid flow, and assuming the bead is neutrally buoyant---a reasonable assumption in the context of swimming---it can be shown that the bead moves as a material point~\cite{mabrouk2024group}. Thus, for each swimmer, we take the velocity of the source (sink) to be equal to the sum of the self-propelled velocity and the velocity induced by the sources and sinks of all other swimmers. A small modification must be made to the velocity of each bead in order to ensure that the length of the dumbbell is constant in time, which is discussed in the ensuing text. 

In this work, we consider a system of $ N $ identical swimmers. We denote by $ \mathbf{x}_{f,i} $ and $ \mathbf{x}_{b,i} $ the position vectors of the source and sink of the $i^{\text{th}}$ swimmer, respectively (Figure~\ref{fig3:SingleSwimmer}). Let $ \mathbf{x}_{c,i} $ be the center position and $ \mathbf{n}_i $ the orientation vector pointing from the sink to the source. We can express $ \mathbf{x}_{f,i} $ and $ \mathbf{x}_{b,i} $ in terms of of $ \mathbf{x}_{c,i} $, $ \mathbf{n}_i $, and $\ell$:
\begin{subequations}
\begin{equation}
\mathbf{x}_{f,i}=\mathbf{x}_{c,i}+\frac{1}{2} \ell \mathbf{n}_i, 
\end{equation}
\begin{equation}
\mathbf{x}_{b,i}=\mathbf{x}_{c,i}-\frac{1}{2} \ell \mathbf{n}_i.
\end{equation}
\label{xcenter}
\end{subequations}
Each swimmer is thus described by six variables: three for the center coordinates and three for the orientation vector. (If desired, the orientation can instead be described by two angular variables.)

The translational and rotational dynamics of each swimmer are given by
\begin{align}
   \frac{\textrm{d} \mathbf{x}_{c,i}}{\textrm{d}t} &= \frac{1}{2} \left( \mathbf{v}_{f,i} + \mathbf{v}_{b,i} \right), \label{translational} \\
   \frac{\textrm{d} \mathbf{n}_i}{\textrm{d}t} &= \frac{1}{\ell} \left( \mathbf{v}_{f,i} - \mathbf{v}_{b,i} + 2\lambda_i \mathbf{n}_i \right) = \frac{1}{\ell} \left\{ \mathbf{v}_{f,i} - \mathbf{v}_{b,i} - \left[ \left( \mathbf{v}_{f,i} - \mathbf{v}_{b,i} \right) \cdot \mathbf{n}_i \right] \mathbf{n}_i \right\}.
\label{RHS normal}
\end{align}
Above, $ \mathbf{v}_{f,i} $ and $ \mathbf{v}_{b,i} $ respectively correspond to the unconstrained velocities of the source and sink of the $i^{\text{th}}$ swimmer due to the self-propelled velocity and the interactions with other swimmers, given by
\begin{subequations}
\begin{equation}
    \mathbf{v}_{f,i} = U \mathbf{n}_i + \frac{\sigma}{4\pi} \sum_{\substack{j=1 \\ j \ne i}}^{N} \left( 
    \frac{\mathbf{x}_{f,i} - \mathbf{x}_{f,j}}{\left\lVert \mathbf{x}_{f,i} - \mathbf{x}_{f,j} \right\rVert^3} 
    - \frac{\mathbf{x}_{f,i} - \mathbf{x}_{b,j}}{\left\lVert \mathbf{x}_{f,i} - \mathbf{x}_{b,j} \right\rVert^3} 
    \right),
\end{equation}
\begin{equation}
    \mathbf{v}_{b,i} = U \mathbf{n}_i  +  \frac{\sigma}{4\pi} \sum_{\substack{j=1 \\ j \ne i}}^{N} \left( 
    \frac{\mathbf{x}_{b,i} - \mathbf{x}_{f,j}}{\left\lVert \mathbf{x}_{b,i} - \mathbf{x}_{f,j} \right\rVert^3} 
    - \frac{\mathbf{x}_{b,i} - \mathbf{x}_{b,j}}{\left\lVert \mathbf{x}_{b,i} - \mathbf{x}_{b,j} \right\rVert^3} 
    \right).
\end{equation}
\label{velpoints}
\end{subequations}
The Lagrange multiplier $\lambda_i$ in \cref{RHS normal} ensures the constraint $ \frac{\textrm{d}\ell}{\textrm{d}t} = 0 $ is satisfied for all swimmers, maintaining the constant distance between the source and sink during motion. It originates from a force that acts on the source and the sink to either push them toward each other or pull them apart, as needed. This force does not produce any net translation or rotation of the $i^{\text{th}}$ swimmer. For more details, see~\cite{mabrouk2024group}. 

We concatenate the position and orientation vectors of all swimmers into a single state vector $\mathbf{X}$ of dimension $6N$. The state of the system is evolved forward in time numerically using the fourth-order Runge-Kutta method. At each step, we re-normalize the orientation vectors $\mathbf{n}_i$ to ensure they have unit length during the simulation.

\section{Triangular configurations}
\label{sec:tri}

We start by investigating the simplest groups with $N > 2$: triangular configurations of swimmers. Five parameters are required to describe the state of our model swimmer: three for its center position and two for its orientation. It therefore seems that 15 parameters are required to describe the state of a trio. The required number of parameters can be reduced to nine by leveraging the continuous translational and rotational symmetries of the system. Even with this reduction, it is infeasible to characterize the dynamics of a trio in such a large parameter space. 

\begin{figure}
    \centering
    \includegraphics[width=0.4\textwidth]{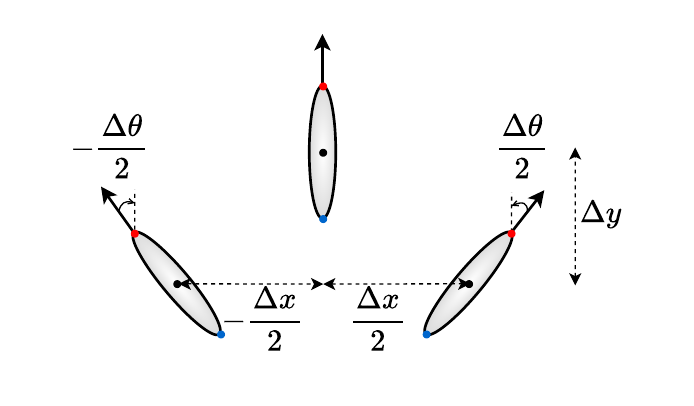}
    \caption{\label{fig:3swconf} Configuration of three swimmers with mirror symmetry about the $y$-axis. One swimmer is fixed at the origin, while the other two swimmers are positioned symmetrically.}
\end{figure}

We therefore focus on a reduced set of configurations that are of the leader-follower type, as sketched in Figure~\ref{fig:3swconf}. Without loss of generality, we fix one swimmer at the origin and orient it in the positive $y$-direction. By making the other two swimmers co-planar with the first and imposing mirror symmetry about the $y$-axis, the configuration of the trio simplifies to being described by two separation parameters, $\Delta x$ and $\Delta y$, and a relative angle between the symmetric pair, $\Delta \theta$. Due to the mirror symmetry, we need only consider initial configurations with $\Delta x > 0$. When $\Delta y > 0$, the trio is arranged as one leader and two followers (which we will also refer to as trailing swimmers). Conversely, $\Delta y < 0$ gives two leaders and one follower. 

Owing to translational symmetry, only the separations $\Delta x$ and $\Delta y$ between the swimmers, not the absolute positions, affect the dynamics. Rotational symmetry further implies that only the relative angle $\Delta \theta$ between the symmetric pair influences the system, as the central swimmer experiences no net rotation or translation in the $x$-direction. The symmetric pair can either enhance or oppose the central swimmer's motion along its axis, while their mirrored hydrodynamic fields preclude lateral motion of the central swimmer.

Several distinct dynamical regimes emerge, depending on the swimmers' initial configuration. The simplest case occurs when $\vert \Delta y \vert \rightarrow \infty$, where the central swimmer moves at its self-propelled speed and the rest of the system effectively reduces to a two-swimmer interaction. In this limit, the two swimmers either end up in deadlock (for $\Delta \theta = -\pi$), collide (for small $\Delta x$), or diverge (for all other parameter values); see \cite{mabrouk2024group} for details. The collision scenario is non-physical in our model, which neglects short-range hydrodynamic effects. 

For finite $\Delta y$, it is useful to think of the symmetric pair as being perturbed by the presence of the central swimmer. All three swimmers now interact, and new dynamical states emerge. We first consider the case where $\Delta y > 0$, corresponding to one leader and two trailing swimmers, before continuing to the case where $\Delta y < 0$, corresponding to two leaders and one trailing swimmer.

\subsection{One leader, two trailing swimmers}

\begin{figure}
    \centering
    \includegraphics[width=0.7\textwidth]{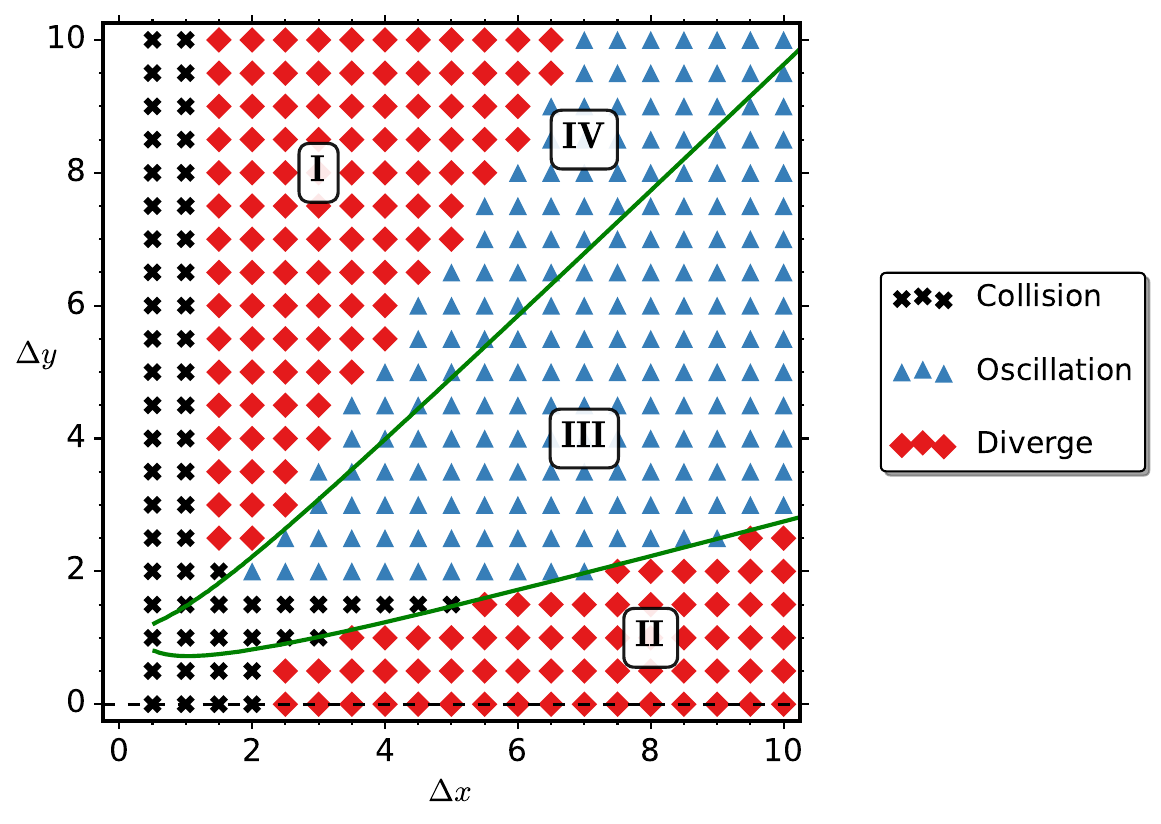}
    \caption{\label{fig:phasediagram1} Phase diagram for $\Delta \theta =0$ for three swimmers, with one leader and two trailing swimmers ($\Delta y > 0$). The green curves correspond to states of the system satisfying $\frac{\textrm{d}\mathbf{n}}{\textrm{d}t} \cdot \mathbf{e}_1 = 0$, where $\mathbf{e}_1$ is the unit vector in the positive $x$-direction. }
\end{figure}

For initial configurations where all three swimmers are aligned ($\Delta \theta= 0$), the phase diagram in Figure~\ref{fig:phasediagram1} summarizes the dynamical states that result. The swimmers either collide, diverge, or oscillate as a cohesive group. We distinguish four regions in the phase diagram. 

Region I consists of tall and narrow configurations of the trio for which the swimmers eventually diverge. The velocity field induced by the leader pulls the trailing swimmers forward towards it, closer towards each other, and tends to rotate them towards each other. The velocity field induced by the trailing swimmers has an opposing effect: it pushes the leader further away, slows down the trailing swimmers, and rotates them away from each other. Because of the tall and narrow configuration, the trailing swimmers are more strongly affected by each other's hydrodynamic fields than the leader's, leading them to rotate away from each other. Furthermore, while the trailing swimmers are pulled forward by the lone leader, the leader is pushed forward by two trailing swimmers, the net effect being that the leader separates from the trailing swimmers. 

\begin{figure}
  \centering
  \subfigure[]{
    \includegraphics[width=0.45\textwidth]{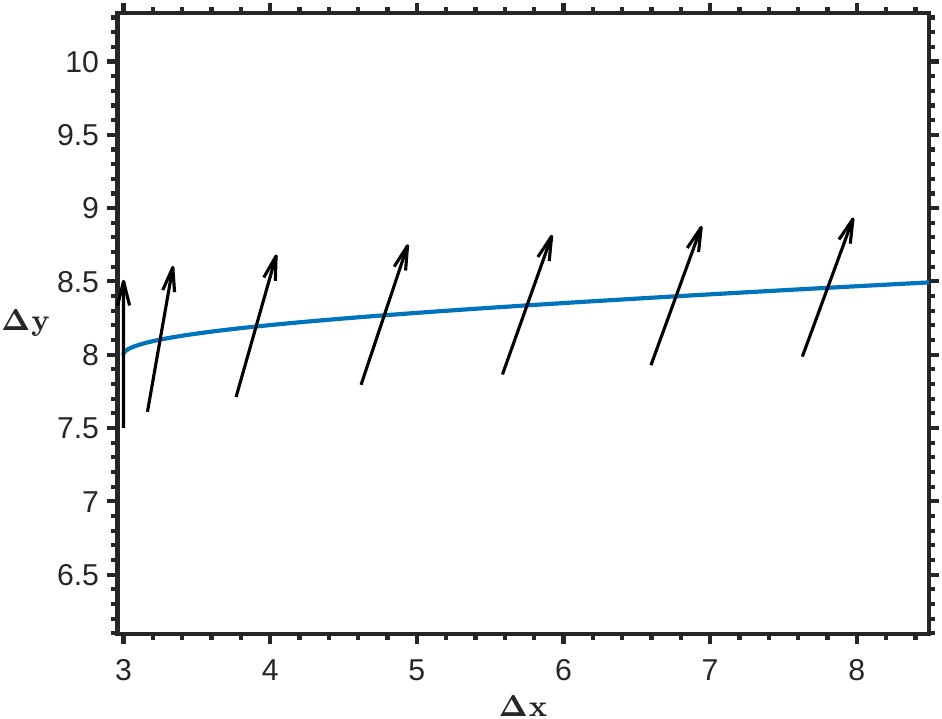}
    \label{fig:p1}
  }
  \subfigure[]{
    \includegraphics[width=0.45\textwidth]{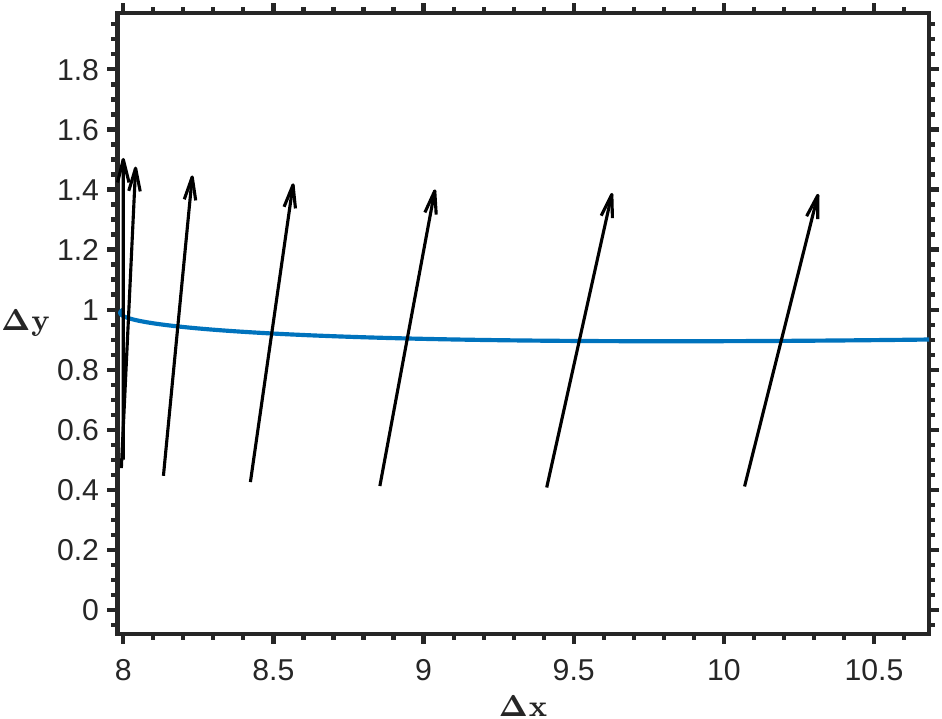}
    \label{fig:p2}
  }
  \subfigure[]{
    \includegraphics[width=0.45\textwidth]{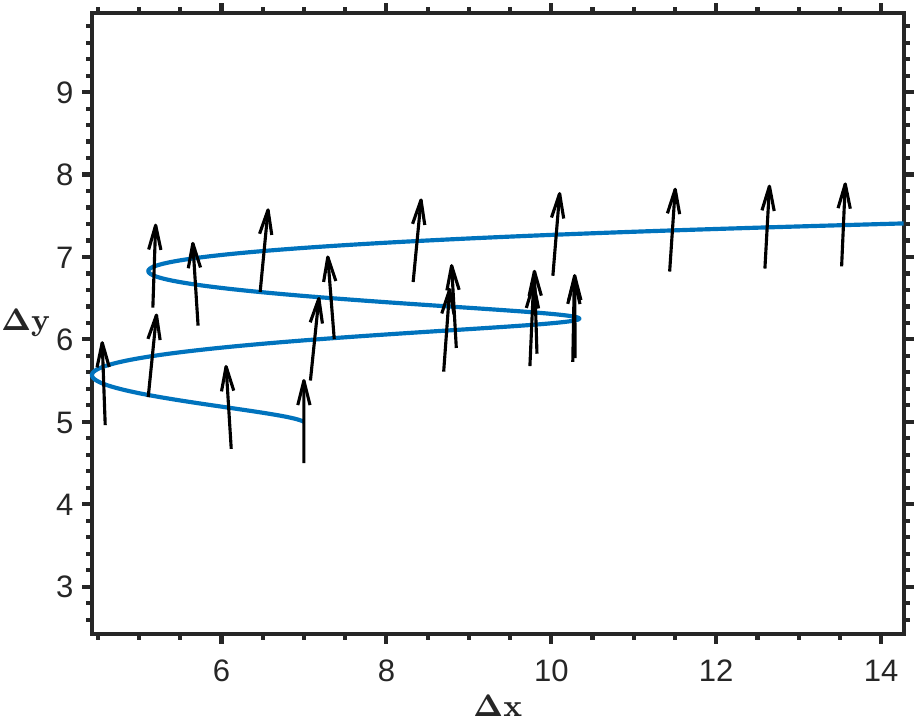}
    \label{fig:p3}
  }
  \subfigure[]{
    \includegraphics[width=0.45\textwidth]{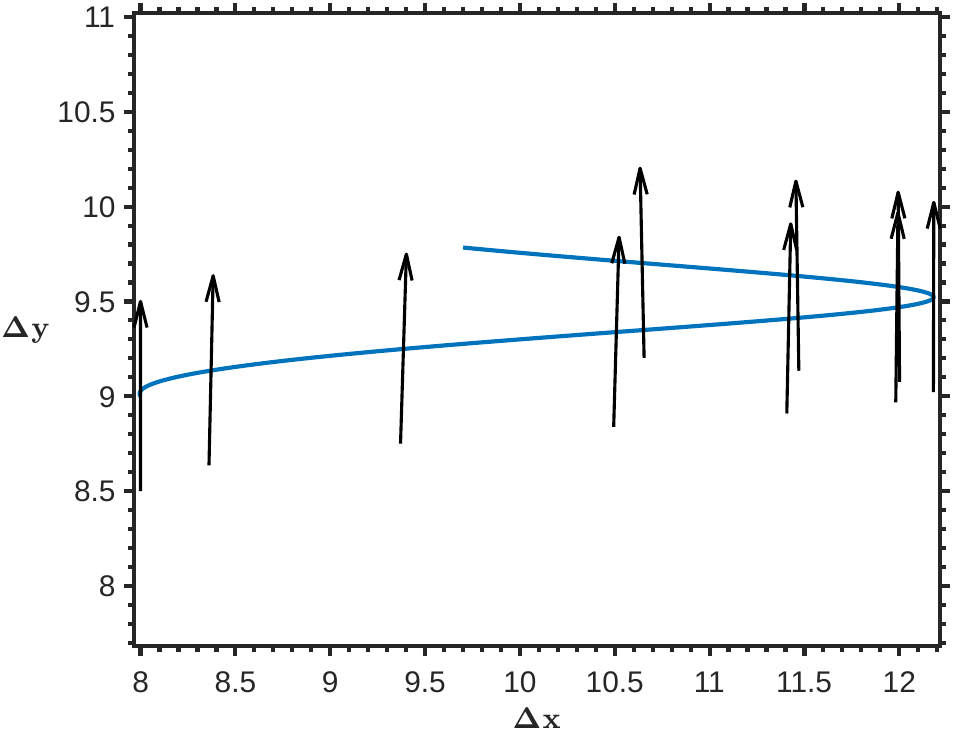}
    \label{fig:p4}
  }
  \caption{\label{fig:Trajectories}Evolution of the relative separations for different initial conditions with $\Delta \theta = 0$.
(a)~Region I: $\Delta x = 3$, $\Delta y = 8$.
(b)~Region II: $\Delta x = 8$, $\Delta y = 1$.
(c)~Region III: $\Delta x = 7$, $\Delta y = 5$.
(d)~Region IV: $\Delta x = 8$, $\Delta y = 9$.
The arrows indicate the vector $(\sin\Delta \theta, \cos\Delta \theta)$, i.e., they are rotated clockwise from the vertical by $\Delta \theta$. N.B.: The plots illustrate the trajectory of the system in the symmetry-reduced space, not the paths taken by the swimmers. }
\end{figure}

A representative case is shown in Figure~\ref{fig:p1}. The vertical separation $\Delta y$ increases monotonically while the lateral separation $\Delta x$ first decreases by a very small amount, reaches a minimum, and then increases. This is accompanied by the two trailing swimmers rapidly rotating away from each other, overshooting their steady-state orientations, before gradually rotating back towards each other and approaching their steady-state orientations where they face away from each other. As they rotate, their self-propelled velocity acquires a lateral component, driving the swimmers outward. As they further separate, the hydrodynamic interactions weaken relative to their self-propelled velocities. Once sufficiently separated, the swimmers are essentially independent of each other, moving forward at their self-propelled speeds, askew, continuing to separate from each other.  

Short and wide configurations also lead the swimmers to diverge from each other, captured by region II in the phase diagram. For such configurations, all the hydrodynamic interactions tend to rotate the trailing swimmers away from each other. The velocity field induced by the leader pulls the trailing swimmers closer to each other. In contrast to the tall and narrow configurations of region I, now the trailing swimmers pull the leader towards them, while the leader pushes the trailing swimmers away from it. Once again, the combined might of the trailing swimmers is stronger, causing the separation in the $y$-direction to initially decrease. A representative case is shown in Figure~\ref{fig:p2}. The lateral and vertical separation of the swimmers initially decreases while the trailing swimmers rotate outward. As they rotate, their self-propelled velocity acquires a lateral component. This lateral component dominates the induced velocities from the other swimmers once they have rotated sufficiently, causing the lateral separation $\Delta x$ to begin to increase. As they continue to separate laterally, the hydrodynamic interactions weaken, and soon the self-propelled velocity dominates the dynamics. At this point, the swimmers move askew from each other along nearly straight lines at their self-propelled speeds, separating vertically and laterally at steady rates. 

Oscillatory dynamics emerge in two adjacent regions (regions III and IV) of the phase diagram. Representative cases are shown in Figures~\ref{fig:p3} and~\ref{fig:p4}, respectively. In both regions, $\Delta y$ increases monotonically while $\Delta x$ exhibits oscillatory behavior on top of a net increase. Similarly, the relative angle $\Delta \theta$ oscillates between positive and negative values. Regions III and IV are distinguished by their initial rotational dynamics, with the trailing swimmers initially rotating inward in region III, whereas they initially rotate outward in region IV. The boundary between the two regions is thus given by the locus of initial configurations for which $\frac{\textrm{d}\mathbf{n}}{\textrm{d}t} \cdot \mathbf{e}_1 = 0$, where $\mathbf{e}_1$ is the unit vector in the positive $x$-direction. This boundary is drawn as a solid green curve in Figure~\ref{fig:phasediagram1}. The two regions are highly intertwined, however. In Figures~\ref{fig:p3} and~\ref{fig:p4}, one can see that $\Delta \theta$ becomes zero at the extremes of the oscillations, and $\frac{\textrm{d}}{\textrm{d}t} \Delta \theta$ changes sign from one extreme to the next. This implies that swimmers whose initial configuration lies in region III move to a configuration lying in region IV, and vice versa. As a result, the swimmers oscillate between the two regions. Physically, when the trailing swimmers are close (i.e., in region IV), they influence each other more strongly than the leader does. The trailing swimmers always rotate each other outward in this region, causing their self-propelled velocities to acquire a lateral component that moves the trailing swimmers apart. Once the trailing swimmers are sufficiently far apart, the influence of the leader becomes relatively stronger. The leader has a tendency to rotate the trailing swimmers inward, which overcomes the outward rotation that the trailing swimmers induce on each other. The trailing swimmers thus rotate inward, reaching region III, and continue rotating to the point where their self-propelled velocities acquire a lateral component that moves them towards each other. The process repeats cyclically. 

The swimmers do not oscillate forever, eventually diverging. The number of oscillations that occur before divergence depends on the initial separation of the swimmers. We have observed anywhere from one oscillation to over 50 oscillations complete before the swimmers diverge. The time between oscillations increases rapidly, in some cases increasing exponentially with the number of oscillations. Since the separation between the leader and the trailing swimmers increases with each cycle, the hydrodynamic influence of the leader on the trailing swimmers decreases from one cycle to the next. As a result, the ability of the leader to rotate the trailing swimmers inward weakens from one cycle to the next. Once the separation has reached a critical value, the leader is unable to rotate the trailing swimmers inward, and they diverge, traveling away at their self-propelled speeds. There is no simple relation for the critical separation since it depends on the maximal relative angle between the trailing swimmers during the oscillations. 

In addition to the boundary separating regions III and IV, there is another locus of configurations for which $\frac{\textrm{d}\mathbf{n}}{\textrm{d}t} \cdot \mathbf{e}_1 = 0$ in Figure~\ref{fig:phasediagram1}. This forms the boundary between the configurations that lead to divergence in region II and those that lead to oscillatory behavior in region III. Such a simple criterion separating the oscillatory configurations of region IV from the diverging configurations of region I does not exist. In both regions I and IV, the trailing swimmers initially rotate outward and then inward. In region I, however, the trailing swimmers attain a larger relative angle $\Delta \theta$, causing them to laterally separate at a faster rate, which causes the hydrodynamic influence of the leader on the trailing swimmers to decay at a faster rate. The leader does not rotate the trailing swimmers inward sufficiently fast, and they diverge from each other. This is analogous to the escape velocity required for an object to escape the gravitational pull of a planet. 

Since the configurations in regions I and II lead to divergence, such trios are not cohesive. While the configurations in regions III and IV lead to hydrodynamic interactions that lock the trio into a dynamically coupled state, their vertical and lateral separation increases on net, and the swimmers eventually diverge. We therefore consider regions III and IV to give rise to semi- (or finite-time) cohesion. For $\Delta \theta \neq 0$, the region of phase space leading to divergence increases in size. 


\subsection{Two leaders, one trailing swimmer}
\label{sec:2lead}

\begin{figure}
    \centering
    \includegraphics[width=0.7\textwidth]{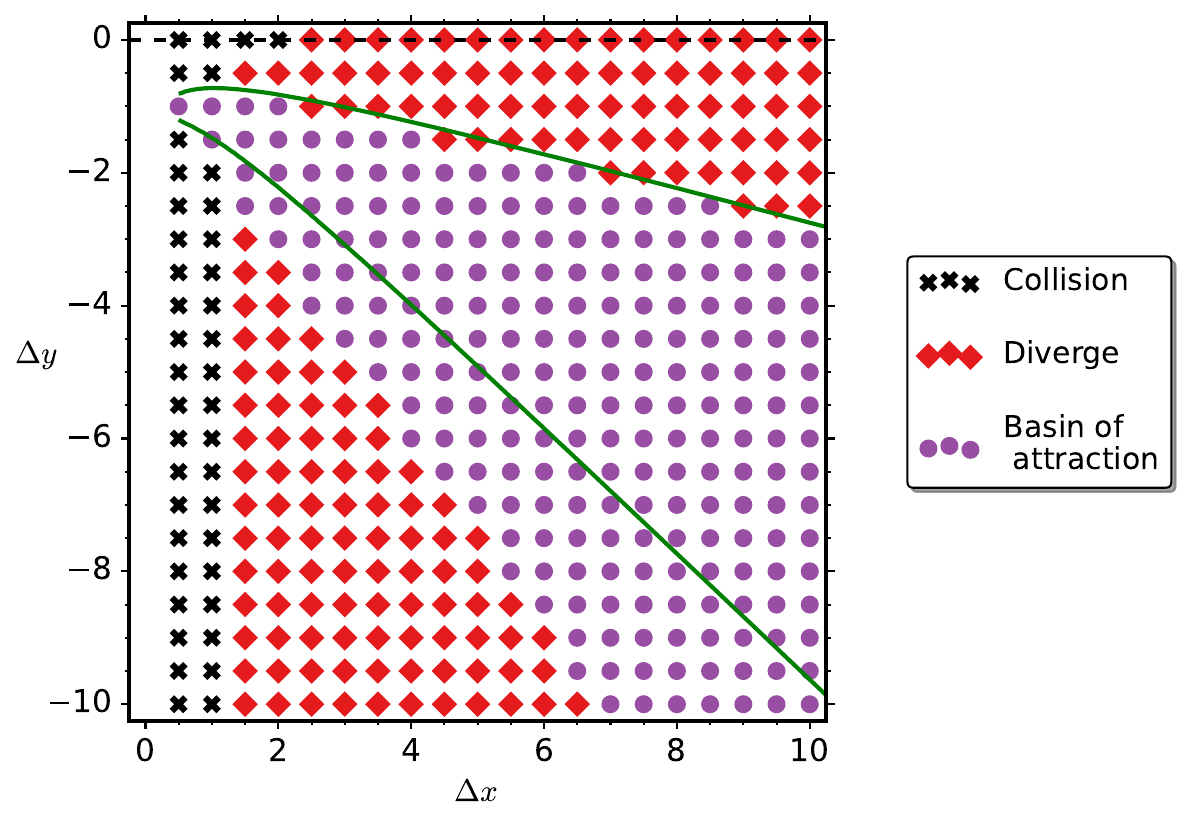}
    \caption{\label{fig:phasediagram2} Phase diagram for $\Delta \theta =0$ for three swimmers, with two leaders and one trailing swimmer ($\Delta y < 0$). The green curves correspond to states of the system satisfying $\frac{\textrm{d}\mathbf{n}}{\textrm{d}t} \cdot \mathbf{e}_1 = 0$, where $\mathbf{e}_1$ is the unit vector in the positive $x$-direction. }
\end{figure}

The phase diagram for a trio with two leaders and one trailing swimmer is shown in Figure~\ref{fig:phasediagram2}. Qualitatively, it is very similar to the phase diagram for one leader with two trailing swimmers: when the swimmers are too close, they collide with each other; and when the aspect ratio of the trio is large or small, the swimmers diverge from each other. 

\begin{figure}
    \centering
    \includegraphics[width=0.4\textwidth]{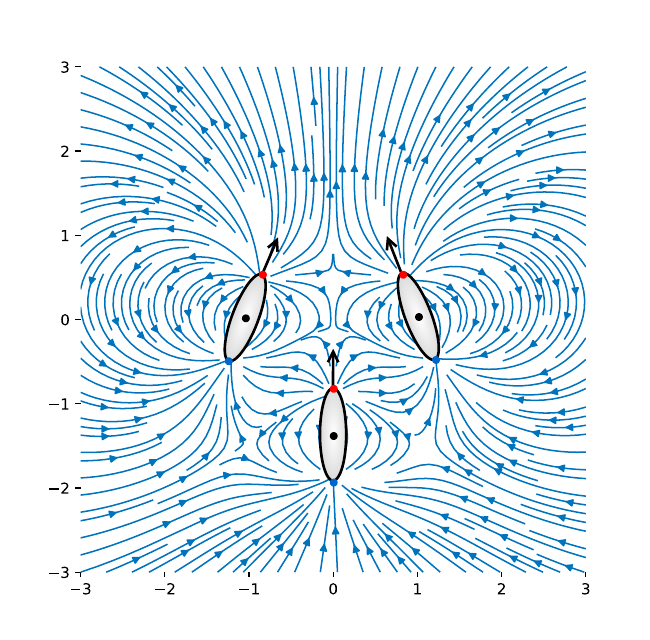}
    \caption{\label{fig:streamlines_fp}Streamlines corresponding to the relative equilibrium. Its configuration is $\Delta x = 2.0779$, $\Delta y = -1.381$, and $\Delta \theta = -0.7247$.}
\end{figure}

The primary difference between the phase diagrams is the emergence of a stable relative equilibrium when there are two leaders and one trailing swimmer. No stable state exists for the case of one leader and two trailing swimmers. The purple circles in Figure~\ref{fig:phasediagram2} show the initial configurations of the trio (with $\Delta \theta = 0$) that converge to the relative equilibrium. The relative equilibrium itself and its induced velocity field are shown in Figure~\ref{fig:streamlines_fp}. The swimmers are configured as an isosceles triangle, with the two leaders facing slightly inward. The group maintains its configuration while moving forward at a speed equal to $0.934 U$, slower than the self-propelled speed of an isolated swimmer. A similar result was demonstrated by \citeauthor{tchieu2012finite} in~\cite{tchieu2012finite}, where a trio of counter-rotating point-vortex pairs evolved toward a stable equilibrium in which the pairs face each other and form an equilateral triangle. Due to their configuration, the vortex pairs do not move. In our case, the configuration is different, and the group has a non-zero speed. 

\begin{figure}
    \centering
    \includegraphics[width=0.6\textwidth]{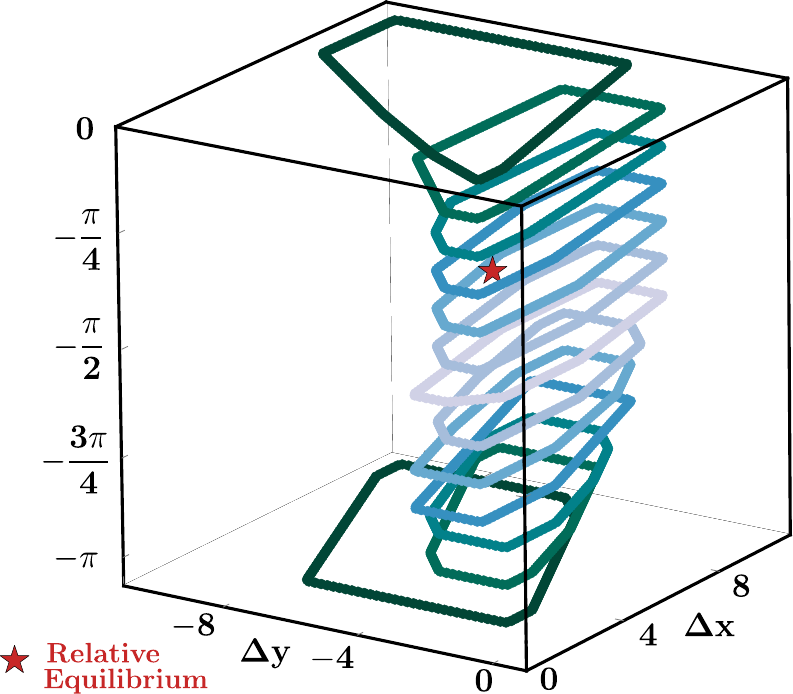}
    \caption{\label{fig:equ3d}Basin of attraction for the relative equilibrium in Figure~\ref{fig:streamlines_fp}, for $\Delta x, \Delta y \leq 10$. Curves are plotted in steps of $\pi/12$ in $\Delta \theta$.}
\end{figure}

A key feature of the relative equilibrium is its robustness to perturbations in $\Delta x$, $\Delta y$, and $\Delta \theta$. The basin of attraction of the relative equilibrium is shown in Figure~\ref{fig:equ3d}, where configurations inside the curves converge to the relative equilibrium. The central swimmer, now trailing the symmetric pair, induces the same rotational dynamics on the symmetric pair as when it was the leader. The translational dynamics that it induces, however, differ. Rather than drawing the pair together, the central swimmer pushes them apart when it is in the trailing position. This difference leads to the emergence of the relative equilibrium. In the equilibrium configuration, the outward rotation that the leaders induce on each other is balanced by the inward rotation induced by the trailing swimmer, and the lateral components of the self-propelled velocities of the leaders are balanced by the laterally outward push induced by the trailing swimmer. 

\section{Diamond configurations}
\label{sec:diamond}

By composing triangular configurations of swimmers, one can create a diamond configuration. In his seminal work, \citeauthor{weihs1973hydromechanics} argued that by arranging themselves into a diamond formation, swimmers stand to expend less energy than in isolation~\cite{weihs1973hydromechanics}. Since then, diamond formations have been among the most studied. Here, we characterize the cohesion of such a formation. 

The stability of diamond lattices of swimmers was previously studied in~\cite{gazzola2016learning}, where it was found that diamond formations are not passively stable. An important distinction between that work and the present work is how the swimmers are modeled. While both utilize a far-field model for the swimmers, \citeauthor{gazzola2016learning} used the model of~\cite{tchieu2012finite}, wherein a swimmer is modeled as a planar pair of point vortices separated by a fixed distance. This produces the same far-field flow as a two-dimensional version of our model, but leads to qualitatively different rotational dynamics due to how the velocity gradient is sampled by the swimmer. As argued in~\cite{mabrouk2024group}, the vortex-pair model is appropriate for short and wide swimmers, whereas the model used in the present is appropriate for more commonly encountered long and narrow swimmers. This key difference may lead to a change in the stability/cohesion of a diamond arrangement of swimmers, as suggested by the results in Section~\ref{sec:2lead}. 


\begin{figure}
    \centering
    \includegraphics[width=0.5\textwidth]{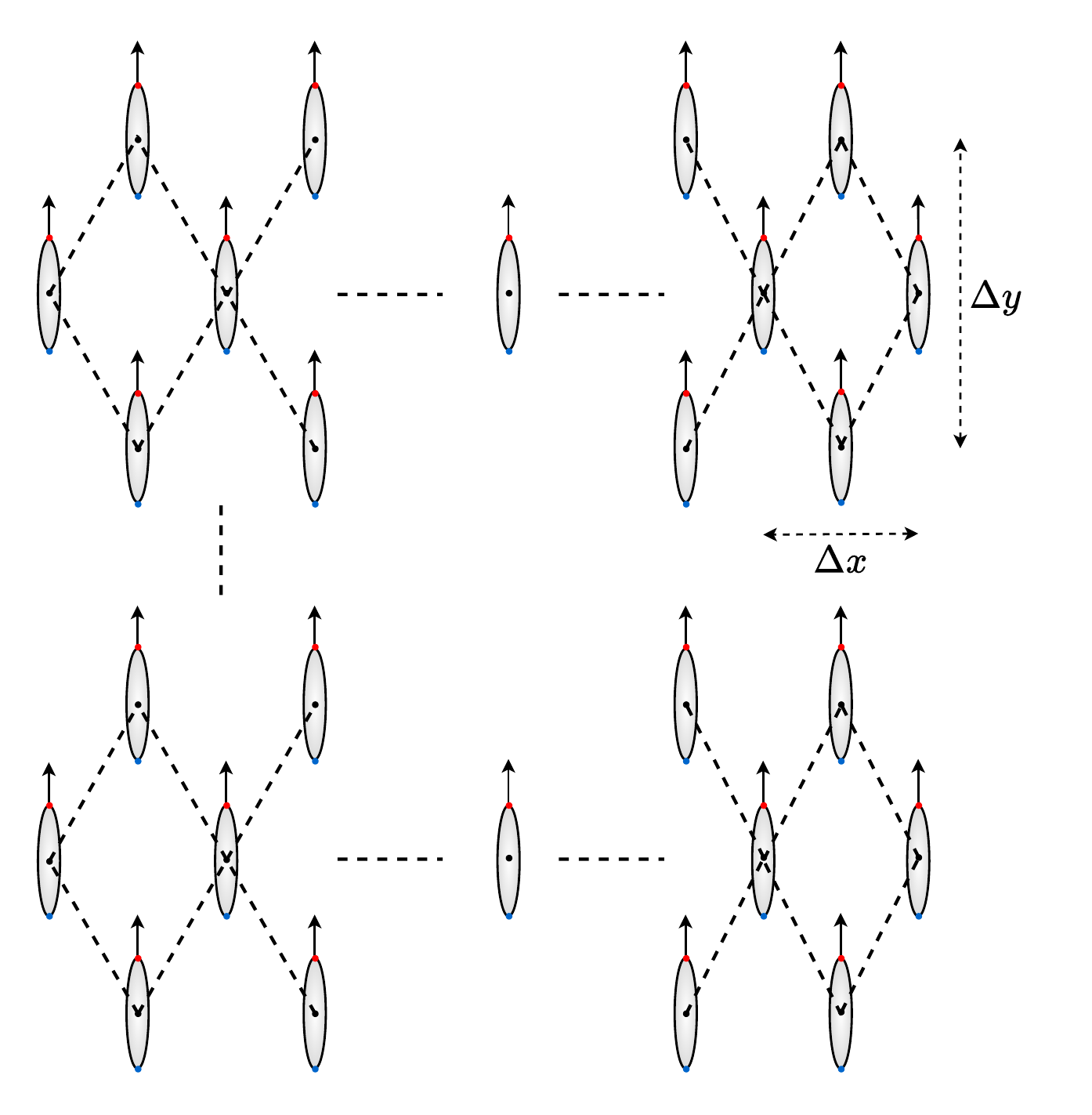}
    \caption{\label{fig:diamondarr} Initial configuration of $N$ swimmers in a planar diamond lattice formation. Arrows indicate identical self-propulsion directions, with uniform lattice spacings $\Delta x$ and $\Delta y$ throughout the school.}
\end{figure}

\begin{figure}
  \centering
  \subfigure[]{
    \includegraphics[width=0.2\textwidth]{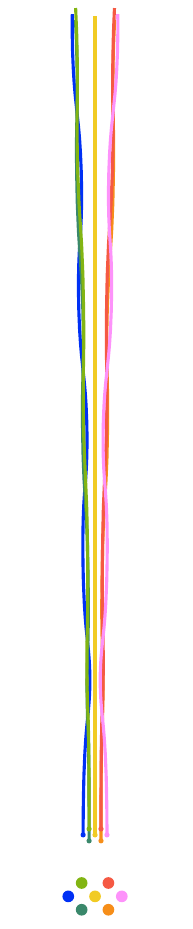}
    \label{fig:7swmotion}
  }
  \subfigure[]{
    \includegraphics[width=0.2\textwidth]{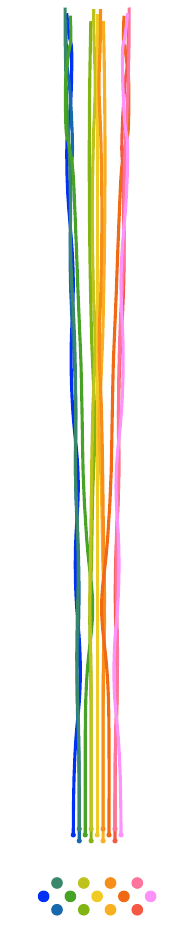}
    \label{fig:13swmotion}
  }
  \subfigure[]{
    \includegraphics[width=0.2\textwidth]{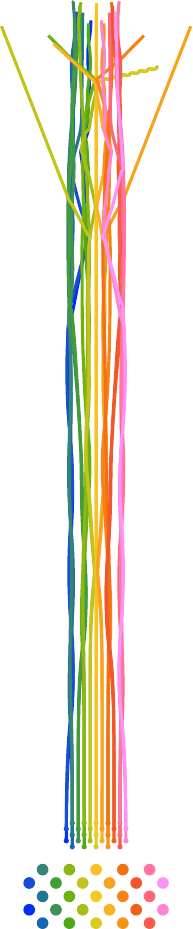}
    \label{fig:27swmotion}
  }
  \caption{\label{fig:Trajectoriesdiam}Planar trajectories of $N$ swimmers for different initial conditions. (a)~$N = 7$, $\Delta x = 16$, $\Delta y = 16$. (b)~$N = 13$, $\Delta x = 16$, $\Delta y = 16$. (c)~$N = 27$, $\Delta x = 16$, $\Delta y = 16$. The patterns on the bottom show the initial arrangements. }
\end{figure}

The diamond arrangement is shown in Figure~\ref{fig:diamondarr}, characterized by the spacings $\Delta x$ and $\Delta y$. We examine three representative cases with group sizes $N = 7$, 13, and 27 swimmers, with fixed spacings of $\Delta x = \Delta y = 16$, whose trajectories are shown in Figure~\ref{fig:Trajectoriesdiam}. All configurations shown have mirror symmetry. 

Several dynamical features emerge. Most obviously, the central swimmers experience neither rotation nor transverse translation. This is a consequence of being along the axis of symmetry. The other swimmers are symmetric about this axis. In fact, the central axis behaves as an impermeable wall since the mirror symmetry is equivalent to using the method of images for potential flows. Consequently, swimmers do not cross the central axis. Surprisingly, the symmetry eventually breaks in Figure~\ref{fig:27swmotion}; we will elaborate on this observation later. 

In the smallest group, the peripheral swimmers separate into subgroups of three swimmers that oscillate in a braided pattern, which often happens to pairs of swimmers~\cite{mabrouk2024group}. The braided trios slowly diverge from each other, and this divergence continues beyond what is plotted, though each subgroup seems to maintain its structure. Although the entire group remains cohesive for only a finite time, the emergent subgroups remain cohesive for much longer. 

The larger diamond formations initially display similar behavior, though in stages. In Figure~\ref{fig:13swmotion}, the outermost swimmers (blue and pink) are the first to form braided trios and move slightly inward. Later, the next layer of swimmers (green and orange) organize into another set of braided trios that move inward. The behavior is similar for the largest group of swimmers in Figure~\ref{fig:27swmotion}, with the outermost swimmers (blue and pink) forming braided quintets, followed by the next layer (green and orange) forming another set of braided quintets, and finally followed by the last layer (light green and light orange) forming braided pairs, all of which move inward. The core of the group is able to sustain nearly straight motion for a longer time than the periphery for the same reason that the central swimmers maintain straight motion: they experience opposing hydrodynamic interactions with swimmers on either side. For example, consider the light green swimmers just left of center in Figure~\ref{fig:27swmotion}. Only the five outermost pink swimmers on the right can cause the light green swimmers to translate laterally or rotate since every other swimmer has an opposite with respect to the light green swimmers that nullifies any induced lateral translation or rotation. Swimmers closer to the periphery experience less such cancellation of hydrodynamic interactions, which is why they veer off straight paths earlier. The blue swimmers on the left, for example, have no leftward neighbors that can oppose the lateral translation and rotation induced by their rightward neighbors, leading the blue swimmers to veer off first. 

Once all braided subgroups have formed from the outside in, they begin to diverge from the inside out. At this point, the dynamics become quite complex, and interactions between subgroups emerge. We observe that swimmers sometimes switch the subgroup they belong to, that subgroups exchange positions, and individual swimmers are ejected from the group. Overall, the group begins to progressively lose cohesion, though not in a catastrophic manner. The eventual loss of cohesion results from a chain of events that starts from the periphery of the group. This suggests that the inner core of a group is passively cohesive, and the cohesion of the overall group can be maintained by actively adjusting or controlling the behavior of the swimmers on the periphery while allowing the inner core of swimmers to passively evolve. 

As we remarked earlier, the largest group in Figure~\ref{fig:27swmotion} loses mirror symmetry. This also happens to the smaller group in Figure~\ref{fig:13swmotion} a bit beyond what has been plotted. Round-off error in our simulations breaks the mirror symmetry; once broken, the dynamics of the system evolve the group further away from a symmetric configuration, which suggests that certain symmetries are not robust to perturbations.

\begin{figure}
  \centering
  \subfigure[]{
    \includegraphics[width=0.25\textwidth]{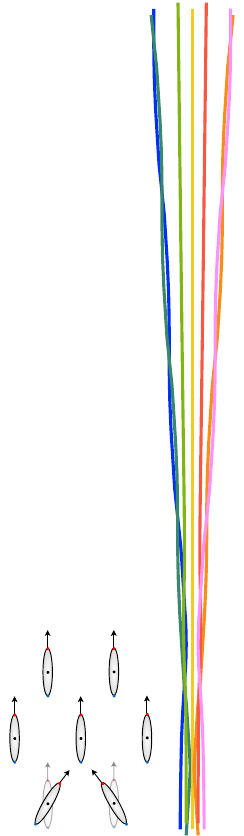}
    \label{fig:per1}
  }
  \subfigure[]{
    \includegraphics[width=0.25\textwidth]{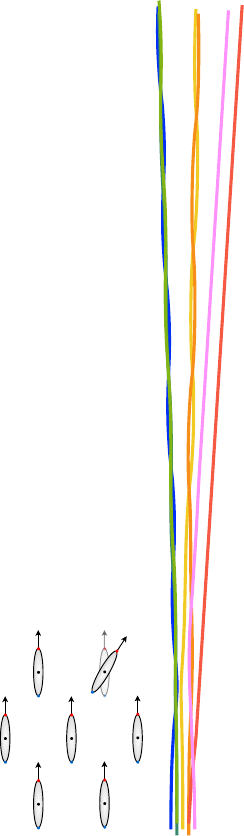}
    \label{fig:per2}
  }
  \caption{\label{fig:petrbdiamond} Perturbed configurations and corresponding trajectories for $N=7$. (a)~Symmetric rotation of two swimmers. (b)~Asymmetric rotation of one swimmer. The perturbed swimmers are initially rotated $\pi/48$ from the vertical.}
\end{figure}

To qualitatively assess the robustness of the diamond arrangement, we consider two types of perturbations for the case $N = 7$. The first is a symmetric perturbation, in which two swimmers are rotated while preserving mirror symmetry. The second is an asymmetric perturbation, where only one swimmer is rotated, thereby breaking the mirror symmetry. The perturbations and resulting trajectories are shown in Figure~\ref{fig:petrbdiamond}, and should be compared to the unperturbed trajectory in Figure~\ref{fig:7swmotion}.

For the symmetrically perturbed case, two swimmers are initially rotated towards each other by $\pi/48$, maintaining the overall mirror symmetry of the arrangement, as illustrated in Figure~\ref{fig:per1}. The swimmers spread out at a faster rate than in the unperturbed case. Further differences arise in the structure of the subgroups, with the peripheral swimmers forming braided pairs instead of trios. Compared to the unperturbed case, one could say that the overall cohesion is somewhat lower, but the dynamics are qualitatively quite similar. 

For the asymmetrically perturbed case, one swimmer is initially rotated outward by $\pi/48$, breaking the mirror symmetry of the arrangement, as shown in Figure~\ref{fig:per2}. The swimmers spread out at a faster rate than in the unperturbed case, but at approximately the same rate as in the symmetrically perturbed case. Subgroups emerge as before; now, the leftmost braided trio from the unperturbed case arises again, and a new central braided pair emerges (though upon closer inspection, the orange swimmer is initially entangled with the red and pink swimmers before finally pairing with the central yellow swimmer). 

Overall, loss of symmetry seems to somewhat lower the cohesion of the group. What is robust to perturbations, however, is the emergence of subgroups that seem to remain cohesive throughout their lifetimes. 

\section{\label{sec4:Milling}Circular configurations}
The last set of configurations that we consider are circular configurations. They are commonly observed, for example, in the milling motion of fish schools, where swimmers follow one another, rotating around an empty core~\cite{cambui2018milling}. Here, we investigate the role that hydrodynamic interactions play in such configurations. We first consider circular configurations for which the motion is in the same plane as the swimmers, and then configurations for which the primary motion is orthogonal to the plane containing the swimmers.

\subsection{In-plane motion}
\begin{figure}
  \centering
  \includegraphics[width=0.45\textwidth]{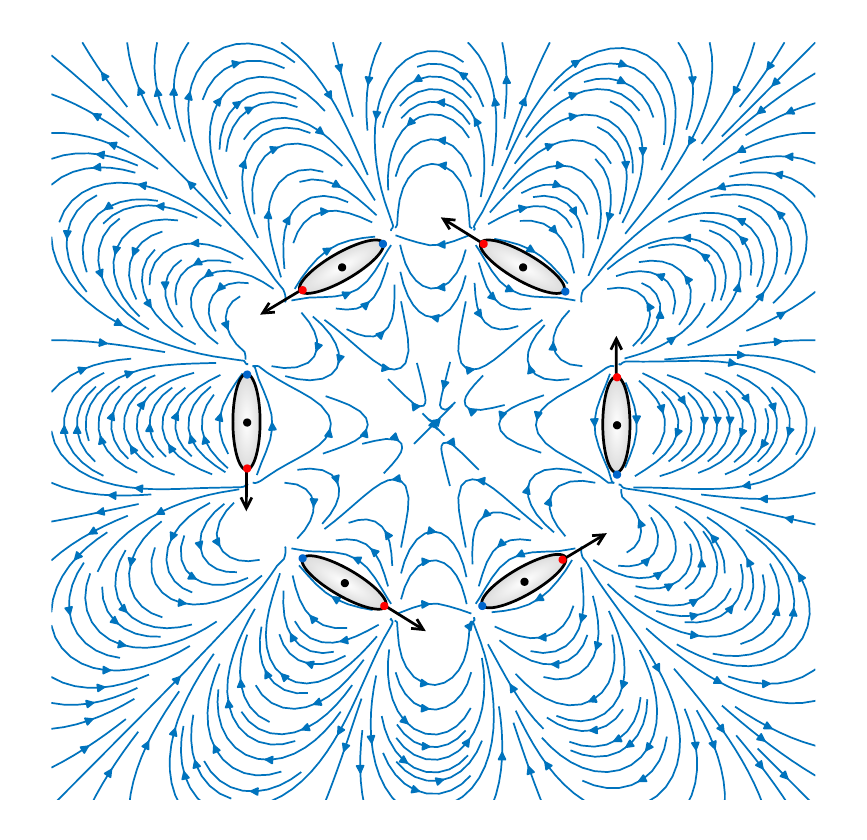}
  \label{fig:stream6}
  \caption{\label{fig:streamlinesmill} Spatial arrangement for swimmers in a circular configuration. Here, $N = 6$, and the streamlines of the induced flow are also shown.}
\end{figure}

We first consider $N$ swimmers that are uniformly distributed along the circumference of a circle of radius $R$. Each swimmer is initially tangent to the circle, as shown in Figure~\ref{fig:streamlinesmill}, which also shows the streamlines of the induced flow. 

\begin{figure}
    \centering
    \includegraphics[width=0.4\textwidth]{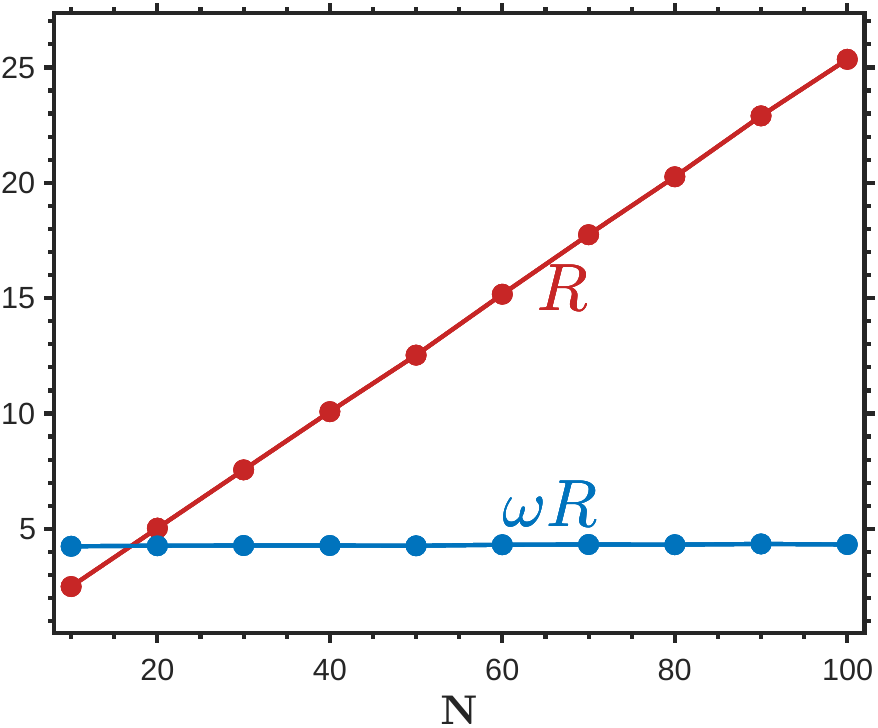}
    \caption{\label{fig:periodic orbits} The radius $R$ and swimmer speed $\omega R$ as functions of group size for hydrodynamic milling configurations. }
\end{figure}

Since milling behavior is commonly observed in nature, we ask whether it can arise purely from hydrodynamic interactions; that is, are there circular configurations of swimmers that maintain their structure indefinitely? With the aid of a Jacobian-free Newton-Krylov solver (see~\cite{willis2019equilibria} for details), for every $N \in [4,100]$, we find an equilibrium circular configuration in which the swimmers chase each other around the circle. Due to its similarity to milling, and its passive hydrodynamic nature, we refer to this behavior as hydrodynamic milling. The behavior is slightly more complicated than simple rotation around a circle; in addition to rotation around the circle, the swimmers' positions undergo small oscillations on a much faster time scale. Dynamically speaking, the circular configurations give rise to relative periodic orbits. For the sake of simplicity, we may ignore the fast oscillations and approximate the motion as simple rotation around a circle. The radius of the equilibrium circular configuration and the speed at which the swimmers move around the circle are shown as functions of $N$ in Figure~\ref{fig:periodic orbits}. The radius is a linear function of $N$, implying that the separation between swimmers remains constant as $N$ increases. The speed of the swimmers is also constant, or, equivalently, the angular rate of rotation of the group, $\omega$, is proportional to $R^{-1}$. 


\begin{figure}
  \centering
  \subfigure[]{
  \includegraphics[width=0.3\textwidth]{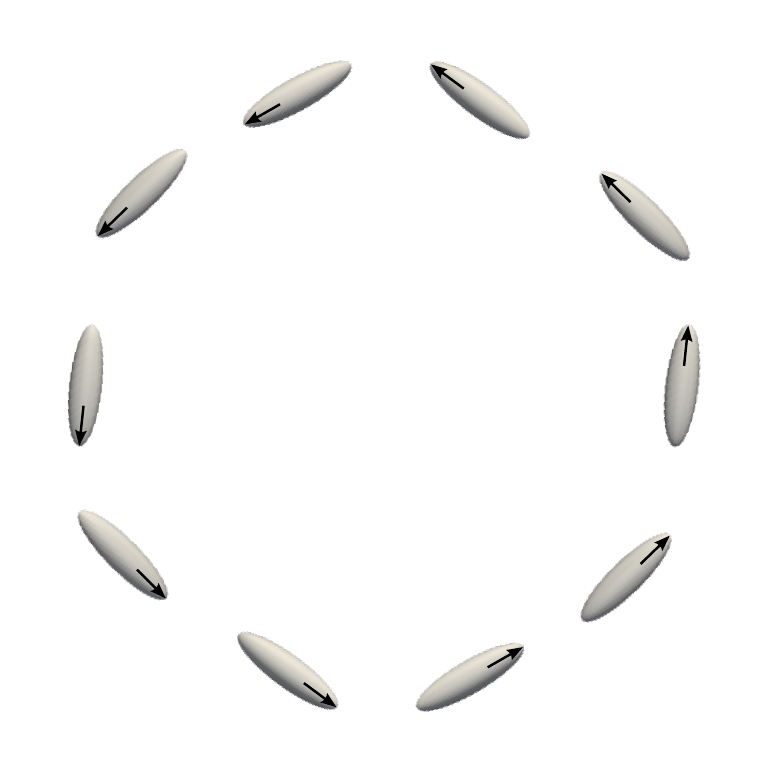}
  \label{fig:mode1}
  }
  \subfigure[]{
    \includegraphics[width=0.3\textwidth]{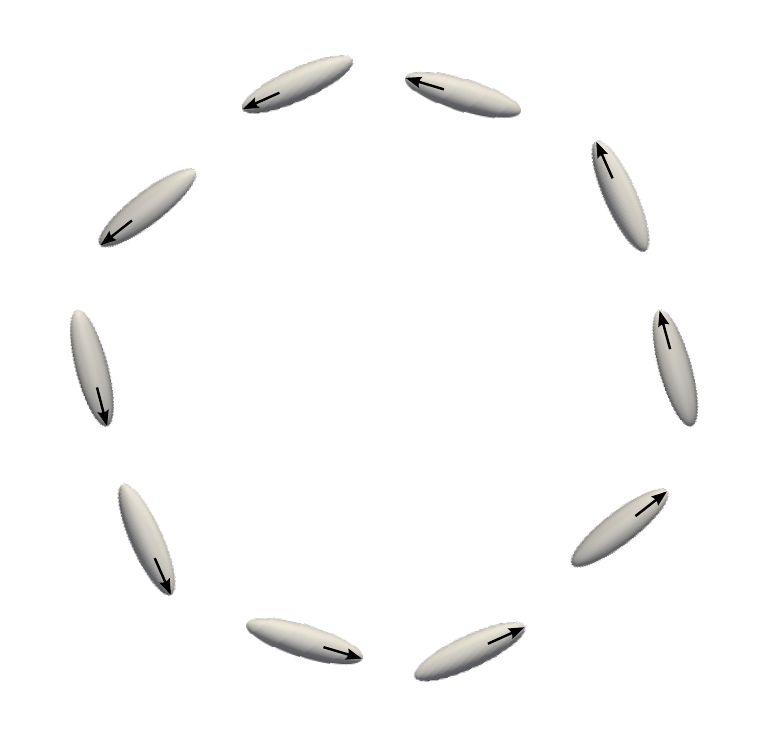}
    \label{fig:mode2}
  }
  \subfigure[]{
    \includegraphics[width=0.3\textwidth]{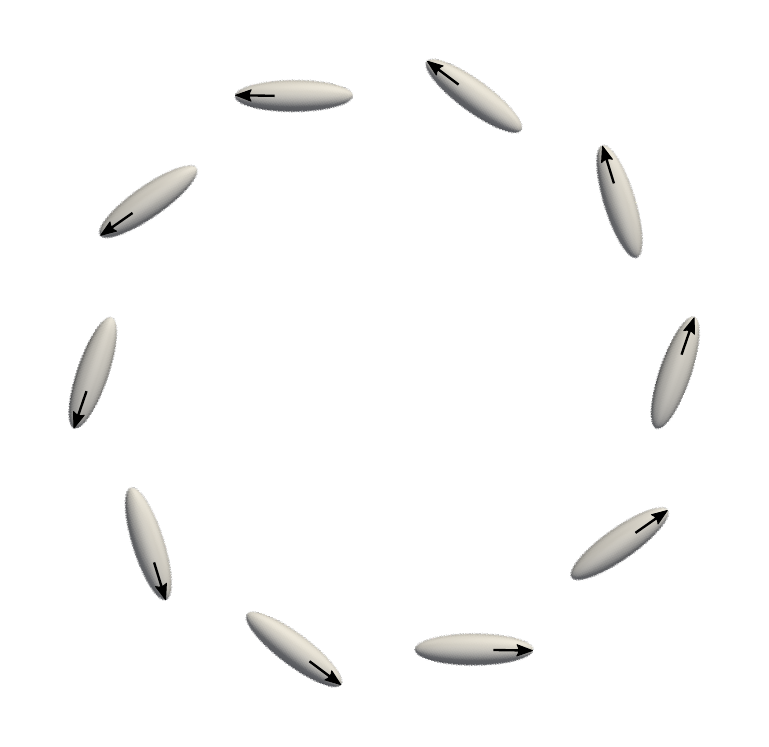}
    \label{fig:mode3}
  }
  \caption{\label{fig:instablemodes} Unstable modes of hydrodynamic milling for $N = 10$. (a)~Mode 1 (asymmetric). (b)~Mode 2 (conjugate to mode 1). (c)~Mode 3 (symmetric).}
\end{figure}

\begin{figure}
  \centering
  \subfigure[]{
  \includegraphics[width=0.3\textwidth]{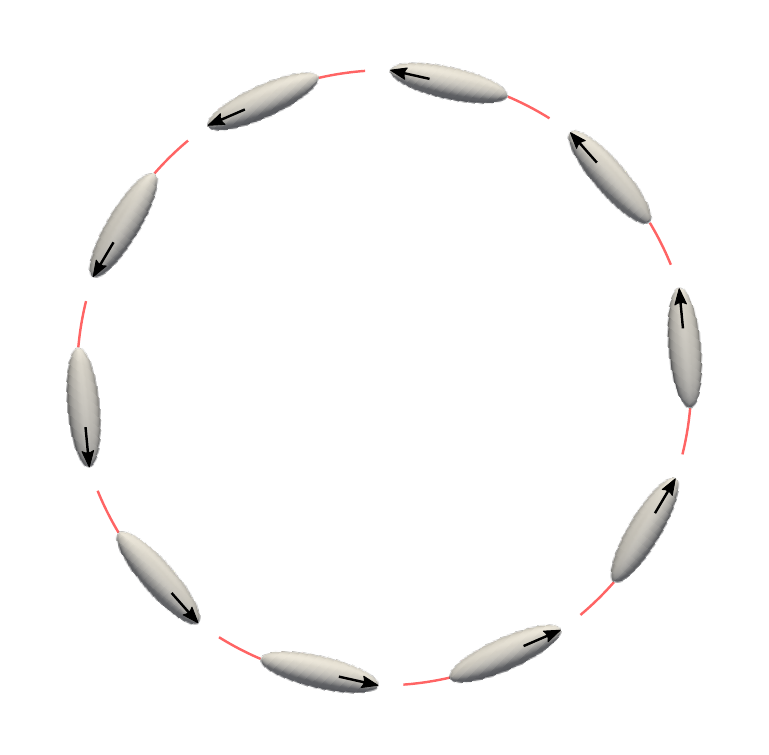}
  \label{fig:periodicmotion}
  }
  \subfigure[]{
    \includegraphics[width=0.3\textwidth]{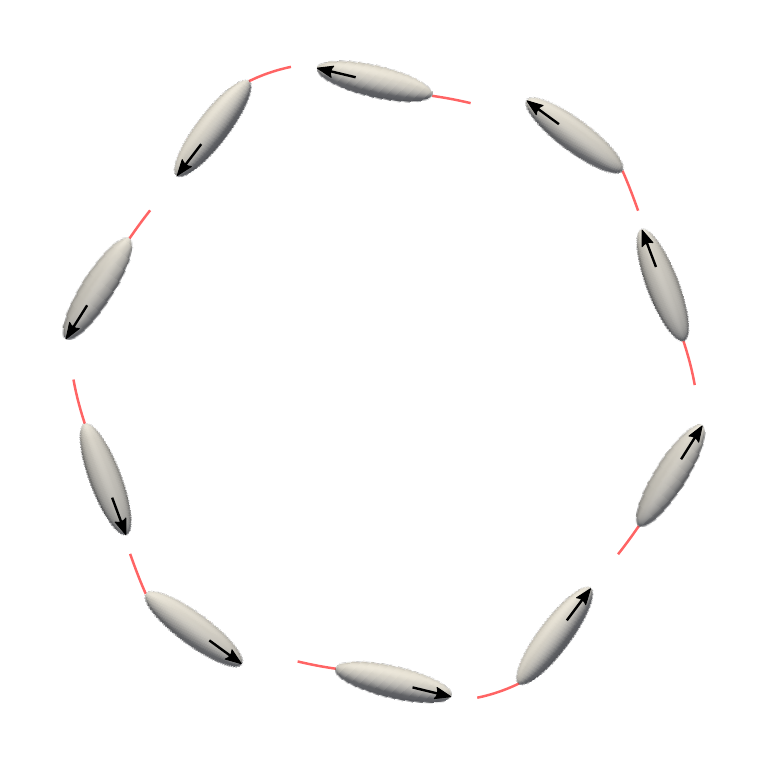}
    \label{fig:onsetofinstability}
  }
  \subfigure[]{
    \includegraphics[width=0.3\textwidth]{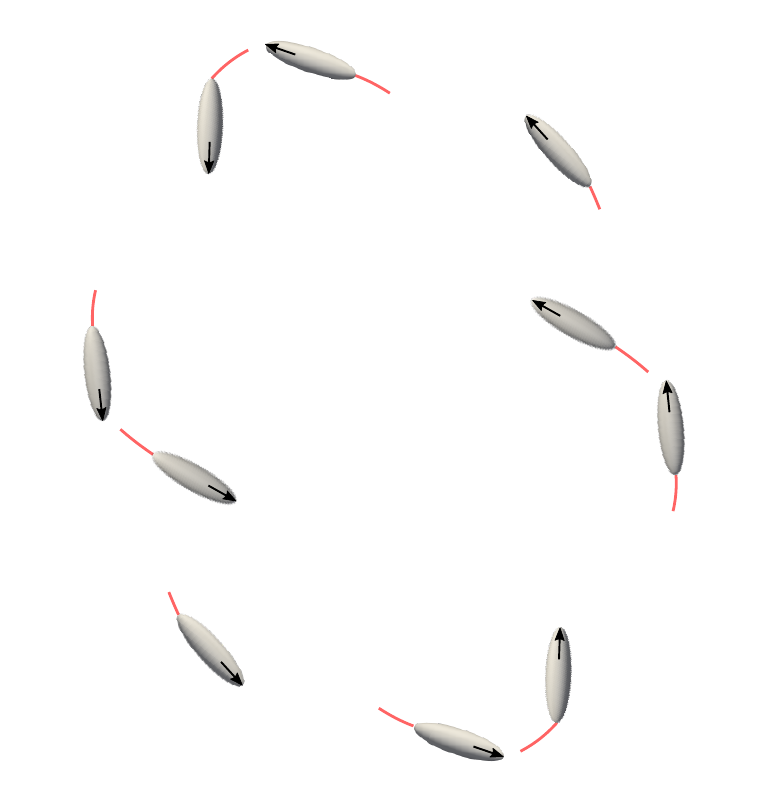}
    \label{fig:splittingintosubgroups}
  }
  \caption{\label{fig:instabilty} Nonlinear evolution of hydrodynamic milling state for $N = 10$. The red comet tails show the trajectories taken in the near past. (a)~Hydrodynamic milling. (b)~Onset of instability. (c)~Splitting into subgroups.}
\end{figure}

To determine the robustness of hydrodynamic milling, we calculate the linear stability of the associated relative periodic orbits. The technical details are given in Appendix~\ref{sec:linstab}. No matter the size of the group, we find that there are always three unstable modes: two asymmetric modes that form a conjugate pair, and a purely real symmetric mode corresponding to uniform rotation of the swimmers. The unstable modes are shown in Figure~\ref{fig:instablemodes}. The conjugate pair always has a faster growth rate than the symmetric mode, which is apparent in the nonlinear simulation shown in Figure~\ref{fig:instabilty}. Beyond the onset of instability, the swimmers split into subgroups before soon colliding. 

\begin{figure}
    \centering
      \subfigure[]{
      \includegraphics[width=0.4\textwidth]{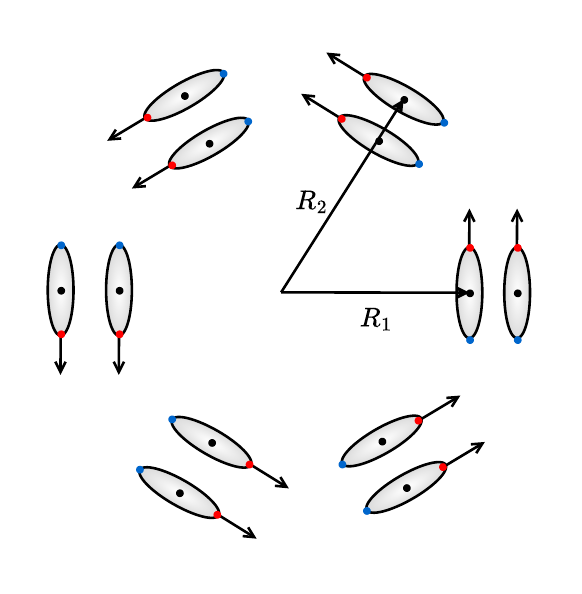}
      \label{fig:configurationnested}
      }
      \subfigure[]{
        \includegraphics[width=0.4\textwidth]{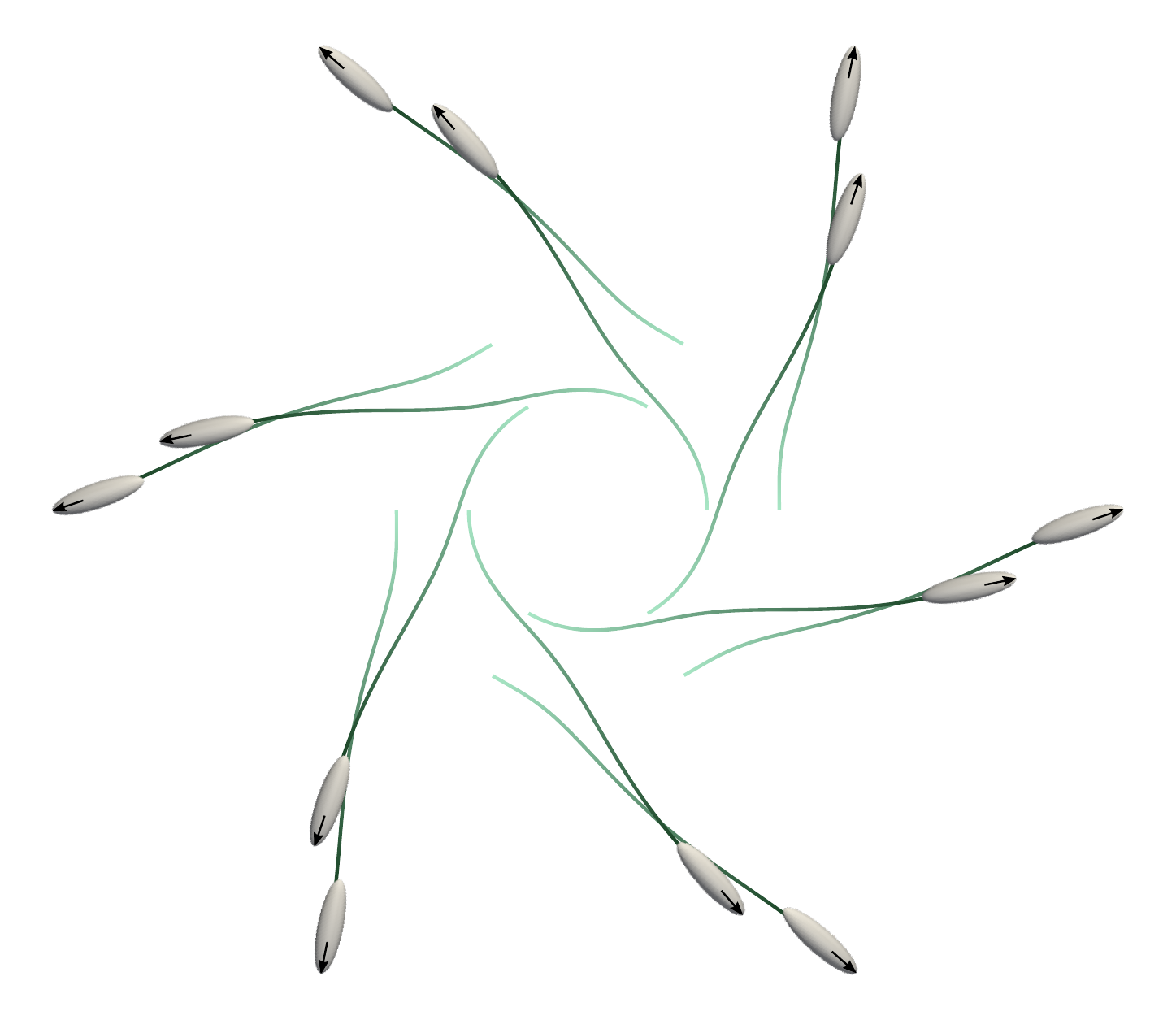}
        \label{fig:traject}
      }
    \caption{\label{fig:nested} (a)~Arrangement of two concentric circles of radii $R_1 = 2.5$ and $R_2 = 4.0$. (b)~Resulting trajectories.}
\end{figure}

More complex circular arrangements ultimately end with the same fate as the hydrodynamic milling state and the diamond configurations. One such example is shown in Figure~\ref{fig:nested}, where twelve swimmers have been initially arranged along two concentric circles. The swimmers destabilize more quickly than in the hydrodynamic milling state, ultimately splitting into pairs that undergo braided motion, as seen previously. Despite the overall group not being cohesive, the subgroups remain cohesive for all later times. 

\subsection{Out-of-plane motion}
\begin{figure}
    \centering
      \subfigure[]{
      \includegraphics[width=0.45\textwidth]{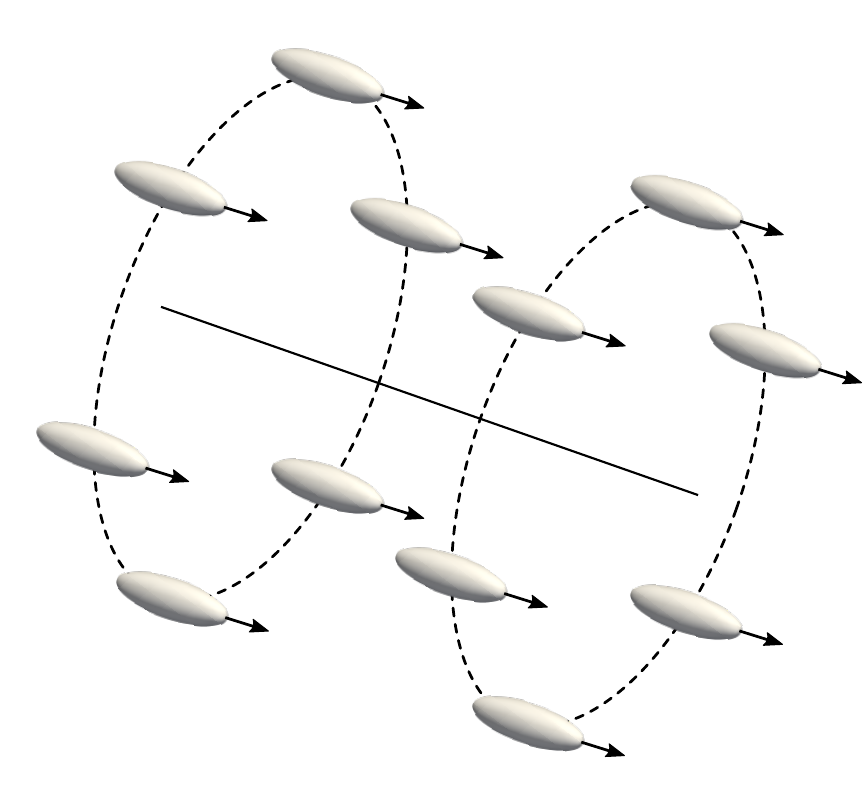}
      \label{fig:configurationrings}
      }
      \subfigure[]{
        \includegraphics[width=0.45\textwidth]{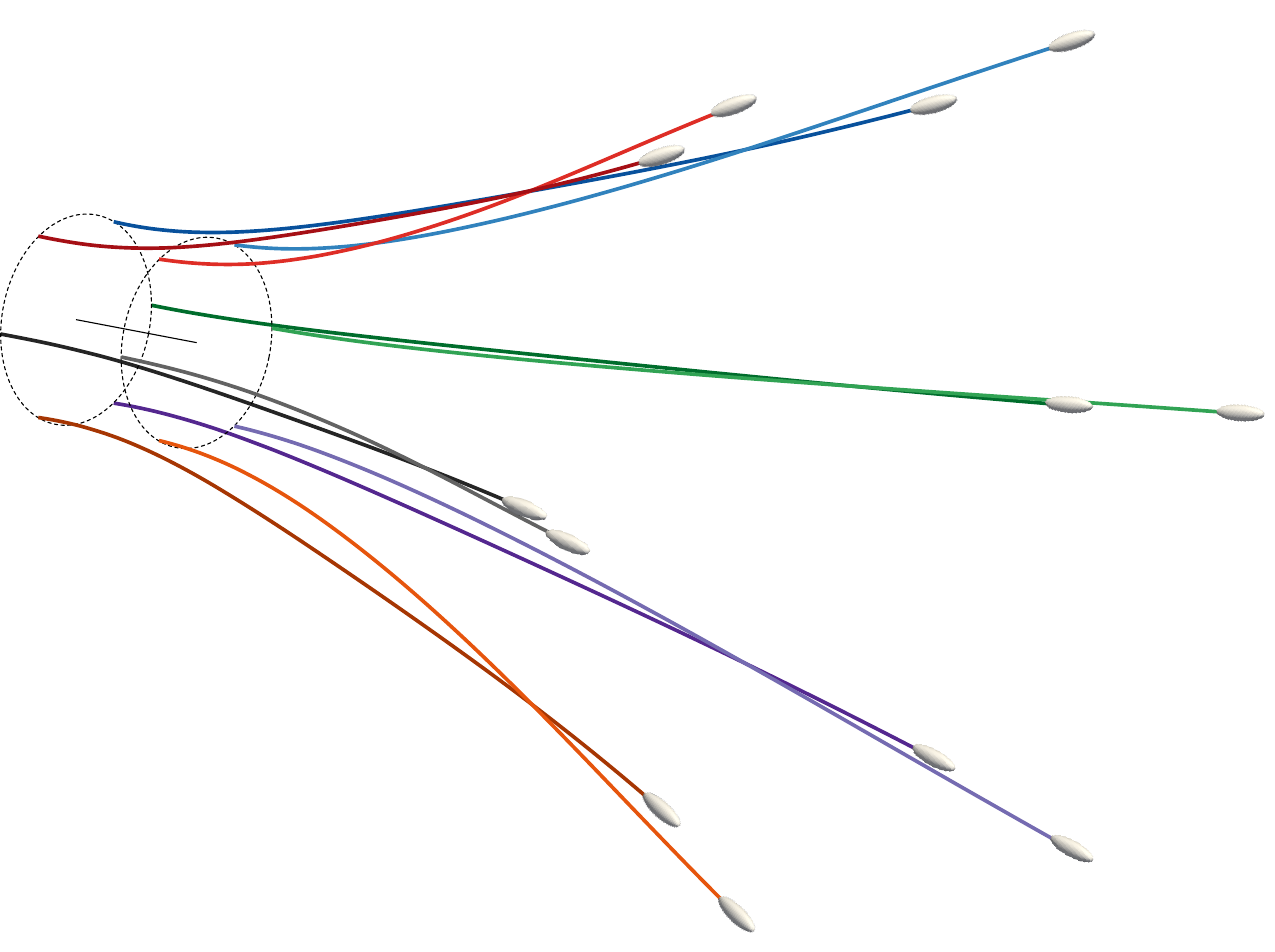}
        \label{fig:trajectrings}
      }
    \caption{\label{fig:rings} (a)~ Swimmers arranged along two coaxial rings $4\ell$. (b)~Resulting trajectories.}
\end{figure}

The final configuration we consider again has the swimmers uniformly distributed along the circumferences of two circles (Figure~\ref{fig:rings}). The circles are co-axial, of the same size, and are offset from each other along their mutual axis. The swimmers differ now in that their orientations are parallel to the mutual axis of the circles rather than tangent to the circles. As was observed for every other configuration with many swimmers, relatively small subgroups quickly emerge. In this case, swimmers from one circle pair up with their counterparts from the other circle. The pairs undergo braided motions and remain cohesive for the rest of time. Even for this fully three-dimensional configuration and motion, the emergence of small and cohesive subgroups remains a robust phenomenon. 

\section{\label{sec6:Conclusion}Summary and conclusions}
In this work, we studied the dynamics and cohesion of freely swimming collectives. Our work was driven by a fundamental question in collective hydrodynamics: can group cohesion be sustained purely through passive hydrodynamic interactions? 

To address this question, we developed a reduced-order potential-flow model for the three-dimensional hydrodynamic interactions of freely moving inertial swimmers. The model represents the far-field flow induced by the motion of long, thin swimmers. In isolation, a model swimmer moves along a straight line at a constant speed. In the presence of a group, each swimmer's motion is modified by the flow induced by all other swimmers, leading to highly non-trivial dynamics. 

Building on previous work in which pairwise interactions were characterized~\cite{mabrouk2024group}, here we studied groups of $N \ge 3$ swimmers. We considered three types of configurations that we view as elemental building blocks of arbitrary groups: triangular, diamond, and circular configurations. 

Swimmers in triangular configurations diverge from each other if the aspect ratio of the triangle is too large or too small. For aspect ratios centered about unity, on the other hand, cohesive groups can emerge. When the configuration consists of one leader and two followers, the group is semi-cohesive in the sense that the swimmers are hydrodynamically locked for a finite time before eventually diverging from each other. Flipping the configuration to two leaders and one follower stabilizes the group: a stable equilibrium configuration emerges, one which has a large basin of attraction. The key factor enabling an equilibrium configuration is that the torques the swimmers induce on each other balance, leading to stable rotational dynamics. 

For larger groups, cohesion is more difficult to maintain. For diamond configurations, subgroups quickly form at the periphery of the group, and then progressively develop inward into the core. Eventually, the entire group splits into subgroups that tend to diverge from each other. Despite the loss of cohesion of the overall group, the subgroups that emerge tend to be cohesive and maintain their structure indefinitely. Additionally, our results are suggestive of how the cohesion of the entire group could be maintained. The ultimate loss of cohesion of the group is precipitated by the changing dynamics of the swimmers at the group's edge, while the core is initially stable. The core destabilizes slowly as the destabilization at the edge diffuses inward. We posit that if the swimmers at the edge are actively controlled, then the core of the group will remain stable under passive dynamics. 

Similar observations are made for circular configurations of swimmers. Interestingly, we discover the existence of hydrodynamic milling states---circular configurations of swimmers that chase each other around a circle while traveling at constant speed. These states reproduce the milling behavior observed in fish schools, though purely through passive hydrodynamic interactions. From the dynamical point of view, the hydrodynamic milling states are relative periodic orbits, and we find that they are always unstable to three modes of instability. The instabilities cause the group to devolve into smaller subgroups that are cohesive, similar to the diamond configuration. Qualitatively similar behavior arises for groups built from multiple circular configurations of swimmers: the overall group is not cohesive, but small cohesive subgroups emerge. 

Returning to our original question of whether group cohesion can be sustained purely through passive hydrodynamic interactions, our results suggest a negative answer. However, there is hope that cohesion can be maintained without resorting to actively controlling every swimmer in the group. Since cohesive subgroups consistently emerge from larger groups, one would only have to control the interactions between subgroups, rather than between all individuals, to create a larger cohesive group. Furthermore, groups seem to progressively destabilize from their periphery inward. If the edge of the group can be controlled, there is hope that the core will remain passively stable. Such a strategy is successfully employed by shepherds and farmers who use dogs to herd livestock; perhaps a similar strategy is possible in aquatic environments as well. \\

This work was supported by the University of Houston Grants to Enhance and Advance Research Program, 000189684.

\appendix
\section{Linear stability analysis of hydrodynamic milling}
\label{sec:linstab}

Let $\mathbf{X} \in \mathbb{R}^{6N}$ be the state of the system, as described at the end of Section~\ref{sec2:Model}. The relative periodic orbits that correspond to hydrodynamic milling are periodic orbits in a frame of reference that rotates with the average rate of rotation of the group. Let $\mathbf{X}_p(t)$ denote the periodic orbit in the rotating frame of reference, so that $\mathbf{X}_p(t) = \mathbf{X}_p(t + T)$, with $T$ the period. One can determine the linear stability of the periodic orbit by performing a Floquet analysis.

Alternatively, one may define a Poincar\'e section, and a point $\mathbf{p}$ along the periodic orbit will be a fixed point of the associated Poincar\'e map $\boldsymbol{\phi}$: $\boldsymbol{\phi}(\mathbf{p}) = \mathbf{p}$. The linear stability of the periodic orbit is equivalent to the linear stability of $\mathbf{p}$ under the discrete dynamics of the Poincar\'e map. We follow this alternative approach. 

To determine the stability of the relative periodic orbit, we require the eigenvalues of the Jacobian of $\boldsymbol{\phi}$ evaluated at $\mathbf{p}$. We estimate the eigenvalues and corresponding eigenvectors using the Arnoldi iteration~\cite{trefethen2022numerical}. The Arnoldi iteration requires the ability to evaluate the product of the Jacobian with any vector. We may estimate such products by using the Taylor expansion of $\boldsymbol{\phi}$, which yields
\begin{equation}
    \boldsymbol{\phi}(\mathbf{p} + \varepsilon \hat{\mathbf{X}}) \approx \boldsymbol{\phi}(\mathbf{p}) + \varepsilon \nabla \boldsymbol{\phi}(\mathbf{p}) \hat{\mathbf{X}} \implies \nabla \boldsymbol{\phi}(\mathbf{p}) \hat{\mathbf{X}} \approx \frac{\boldsymbol{\phi}(\mathbf{p} + \varepsilon \hat{\mathbf{X}}) - \boldsymbol{\phi}(\mathbf{p})}{\varepsilon}, \quad \varepsilon \ll 1.
\end{equation}
Thus, the Jacobian-vector product $\nabla \boldsymbol{\phi}(\mathbf{p}) \hat{\mathbf{X}}$ can be approximated by simply evaluating $\boldsymbol{\phi}$. By definition, $\boldsymbol{\phi}(\mathbf{p}) = \mathbf{p}$. To get $\boldsymbol{\phi}(\mathbf{p} + \varepsilon \hat{\mathbf{X}})$, we perturb $\mathbf{p}$ on the Poincar\'e section by $\varepsilon \hat{\mathbf{X}}$ and use our nonlinear solver to evolve the solver forward in time until the state intersects the Poincar\'e section.

\bibliography{apssamp.bib}
\end{document}